\documentclass[12pt]{article}

\usepackage{amsfonts,amsmath,amsxtra}

\usepackage{graphicx}

\newcommand{\beq}[1]{\begin{equation}\label{#1}}
\newcommand\eeq {\end{equation}}
\newcommand\bqa {\begin{eqnarray}}
\newcommand\eqa {\end{eqnarray}}
\newcommand\pr {\partial}

\newcommand{\eq}[1]{eq.(\ref{#1})\ }
\newcommand{\bear}{\begin{array}}
\newcommand{\enar}{\end{array}}

\newcommand{\C}{\mathbb{C}}

\begin{document}
\def\t{\theta}
\def\T{\Theta}
\def\w{\omega}
\def\ov{\overline}
\def\a{\alpha}
\def\b{\beta}
\def\g{\gamma}
\def\s{\sigma}
\def\l{\lambda}
\def\wt{\widetilde}


\hfill{ITEP-TH-19/05}

\vspace{5mm}

\centerline{\bf \Large Towards the Theory of Non--Abelian Tensor
Fields I}

\vspace{5mm}

\centerline{{\bf Emil T.Akhmedov
}\footnote{email:{akhmedov@itep.ru}}}

\vspace{5mm}

\centerline{117218, Moscow, B.Cheremushkinskaya, 25, ITEP, Russia}




\begin{abstract}
We present a triangulation--independent area--ordering
prescription which naturally generalizes the well known path
ordering one. For such a prescription it is natural that the
two--form ``connection'' should carry three ``color'' indices
rather than two as it is in the case of the ordinary one--form
gauge connection. To define the prescription in question we have
to define how to {\it exponentiate} a matrix with three indices.
The definition uses the fusion rule structure constants.
\end{abstract}

\vspace{3mm}

\section{Introduction}

Among the questions standing in front of the modern mathematical
physics there are two which are of interest for us in this note:

\begin{itemize}

\item What is the theory of non--Abelian tensor fields?

\item How to define ``multiple--time'' Hamiltonian formalism for
non--point--like objects?

\end{itemize}

We argue in this note that these two questions are related to each
other.

  To define the theory of the two--tensor field we would like to
understand its nature. We think that the appropriate point of view
on the tensor field is to consider it as a connection on an
unusual ``fiber bundle''. Or, better to say, we present a somewhat
different way of looking at known type of fiber bundles, where the
two--tensor field acquires its natural place. The base of this
bundle is the loop space ${\cal L} X$ of a finite--dimensional
space $X$. A ``point'' of the space ${\cal L} X$ is the loop
$\gamma$ on the space $X$. A ``path'' connecting two ``points''
$\gamma_1$ and $\gamma_2$ of the base ${\cal L} X$ is the surface
$\Sigma(\gamma_1,\, \gamma_2)$ inside the space $X$. The surface
$\Sigma(\gamma_1,\, \gamma_2)$ has cylindrical topology with the
two end--loops $\gamma_1$ and $\gamma_2$.

Hence, the connection on this bundle $\hat{B}$ is a two--form
which has to be integrated over the surface $\Sigma$. This is
similar to the standard situation with the string two--tensor
$B$--field. However, unlike that situation we would like to
consider such fibers $V$ (sitting at each of the points of the
loops $\gamma_1$ and $\gamma_2$), which have dimensionality bigger
than one. Hence, the $\hat{B}$ field in this case carries
``color'' indices. As well the corresponding ``holonomy matrix''
of the $\hat{B}$ connection over the ``path'' $\Sigma$ has
continuous number of indices distributed on the end--loops
$\gamma$'s. To define the holonomy of the $\hat{B}$ field we have
to define the area--ordering prescription.

We are not trying to formalize those concepts rather we would like
to present an explicit construction of the objects listed in the
previous paragraph. To define the area--ordering we consider a
triangulation $\tilde{\Sigma}$ of the Riemann surface $\Sigma$. In
this case the area--ordering is obtained via gluing over the whole
simplicial surface {\it exponents} of $\hat{B}$'s assigned to each
simplex. But from this picture it is obvious that the {\it
exponents} should carry three indices in accordance with three
wedges of each simplex --- triangle. The question is what is the
{\it exponent} which has three indices\footnote{See
\cite{Chernyakov} on various features of cubic matrices and on
their natural multiplication rules.}? To move further we suppose
that this is a new object
--- a function of the matrix $\hat{B}$ which has three indices as
well. We are going to define this function in this note. To define
it we have to obey the main properties necessary to make the
construction in question meaningful. The construction is
meaningful if the prescription of the area--ordering does not
depend on the way the continuum limit (from $\tilde{\Sigma}$ to
$\Sigma$) is taken: We refer to this fact as the triangulation
independence of the definition of the area--ordering.

Thus, the main difficulty for this definition is the absence of
the understanding of how to {\it exponentiate} matrices with three
indices. In this note we give a definition of such an {\it
exponent}. To give an idea of this {\it exponent} and of the
``triangulation independence'', let us recall that the exponent of
a matrix ${A_i}^j$ with two indices has the following main
feature:

\bqa {\left(e^{t_1 \, \hat{A}}\right)_i}^j \, {\left(e^{t_2 \,
\hat{A}}\right)_j}^k = {\left(e^{\left(t_1 + t_2\right)\,
\hat{A}}\right)_i}^k \eqa for any two numbers $t_1$ and $t_2$.
This equation defines the exponent unambiguously. In fact, from
this equation one can derive the differential equation for the
exponent.

 The {\it exponent} of a matrix $B_{ijk}$ with three indices
can have \underline{any} number of external indices (not only zero
or two as the exponent of a matrix with two indices). As the
result it obeys many different identities following from the
conditions of the triangulation independence. Furthermore, the
function of the three--linear form $B$ in question is not really
an exponent in commonly accepted sense, but we refer to it as
``{\it exponent}'' for the reason it obeys such conditions of
triangulation independence.

\begin{figure}
\begin{center}
\includegraphics[scale=0.5]{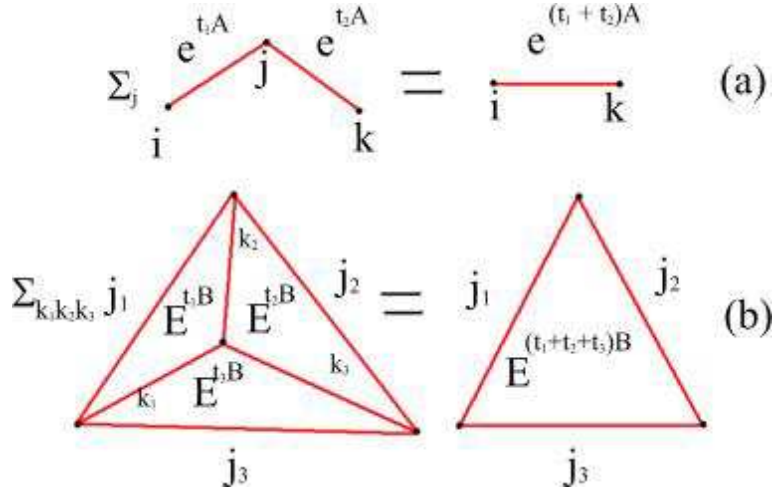}\caption{\footnotesize Main properties of the
exponents.}\label{fig0}
\end{center}
\end{figure}

The most beautiful identity obeyed by the {\it exponent} in
question is as follows\footnote{The generalization of these
equations to higher dimensions is obvious: Through the barycentric
decomposition of the multi--dimensional simplicies.}:

\bqa {\left(E^{t_1 \, \hat{B}}\right)_{j_1\, k_2}}^{k_1}
\,{\left(E^{t_2 \, \hat{B}}\right)_{j_2\, k_3}}^{k_2}\,
{\left(E^{t_3 \, \hat{B}}\right)_{j_3\, k_1}}^{k_3} =
\left(E^{\left(t_1 + t_2 + t_3\right)\, \hat{B}}\right)_{j_1\,
j_2\, j_3},\label{baric}\eqa which is shown graphically in the
fig. \ref{fig0}. This equation, however, does not define
unambiguously the {\it exponent} of the matrix $\hat{B}$ with
three indices.
However, it is this point where the formalism which we develop can
help in establishing ``two--time'' Hamiltonian formalism. We
briefly discuss this point in concluding section. A more complete
discussion will be given in another publication.

It is worth mentioning at this point that the area--ordering can
be defined via the standard exponent for a two--tensor form with
two ``color'' indices. But such a definition demands as well the
presence of an additional one--form gauge connection
\cite{Dolotin:1999hk}. Hence, if we would like to deal with the
two--tensor field only we have to follow our line of reasoning.

The organization of the paper is as follows. In the section 2 we
present our prescription of the area--ordering. In the section 3
we define how to {\it exponentiate} matrices with three indices.
In the section 4 we discuss the properties of the area--ordering
and of the ``fiber bundle'' in question. In this section as well
we present other definitions of the {\it exponent}. In the section
5 we present some explicit examples of the $I$ and $\kappa$
matrices to be defined in the main body of the text. We end up
with conclusions and brief discussion of some future direction
(such as development of the ``two--time'' Hamiltonian formalism).
Appendix contains some necessary explicit calculations.

\section{Area--ordering and triangulations}

\begin{figure}
\begin{center}
\includegraphics[scale=0.7]{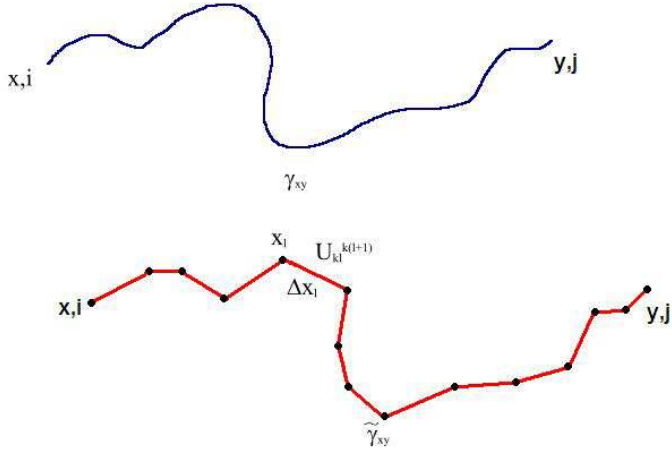}\caption{\footnotesize $\tilde{\gamma}_{xy}$
is the discretization of $\gamma_{xy}$. On each wedge of
$\tilde{\gamma}$ there is ${U_{k_l}}^{k_{(l+1)}}$.}\label{fig1}
\end{center}
\end{figure}

To set the notations and to present the idea of our argument let
us describe briefly how one obtains the path ordering. Consider
the base space $X$ of a vector bundle with $N$--dimensional fibers
$V$. We would like to find the holonomy matrix, corresponding to a
path $\gamma_{xy}$, which relates $N$--vectors in fibers over two
points (say $x$ and $y$) of the $X$ space. We give here a somewhat
weird way of defining the path ordered exponent, which, however,
is easy to generalize to the two--dimensional case. To define the
holonomy matrix we approximate the path $\gamma_{xy}$ by a broken
line $\tilde{\gamma}_{xy}$, consisting of a collection of small
straight lines (see fig. \ref{fig1}).

The holonomy in question is given by:

\bqa {U\left(\tilde{\gamma}_{xy}\right)_i}^j \equiv \sum^N_{k_1,
k_2, \dots, k_L} {U\left(x,\Delta_1 x\right)_i}^{k_1} \,
{U\left(x_1, \Delta_2 x \right)_{k_1}}^{k_2} \dots {U(y, \Delta_L
x)_{k_L}}^j,\label{hol}\eqa where $L$ is the number of wedges of
the broken line; $x_1 = x + \Delta_1 x$, etc. and $\Delta_l x$ is
the $l$-th wedge of the broken line; each $U$ in \eq{hol} is an
operator $\hat{U}: V\to V$. From now on small Latin letters
($i,j,k$ etc.) represent ``color'' indices running from $1$ to
$N$.

To obtain the holonomy for the path $\gamma_{xy}$ itself we have
to take the continuum limit $L\to\infty$ and $\left|\Delta_l
x\right|\to 0$, $l=1, ..., L$. This definition does not depend on
the one--dimensional triangulation, i.e. on the concrete choice of
the sequence of $\tilde{\gamma}_{xy}$'s in approaching
$\gamma_{xy}$ as $L\to\infty$. This is true due to the fact that
each $U$ in the product (\ref{hol}) can be represented as the
exponent of an element of the algebra --- the connection of the
vector bundle. The definition of the exponent is (do not confuse
it with the path ordered exponent):

\bqa {U_i}^j = {\left(e^{\hat{A}}\right)_i}^j \equiv
\lim_{M\to\infty} {\left(\prod_{\rm graph}^M \left[1 +
\frac{\hat{A}}{M}\right]\right)_i}^j \equiv \nonumber
\\ \equiv \lim_{M\to\infty} \left({\delta_i}^{j_1} +
\frac{{A_{i}}^{j_1}}{M}\right)\left({\delta_{j_1}}^{j_2} +
\frac{{A_{j_1}}^{j_2}}{M}\right) \dots \left({\delta_{j_M}}^{j} +
\frac{{A_{j_M}}^{j}}{M}\right) = \nonumber
\\ \lim_{M\to\infty}\left[{\left(\prod_{\rm graph}^M
\delta\right)_i}^j + \frac{1}{M}\, \sum_{a=1}^M {A_{k}}^m(a)
\left(\prod_{{\rm graph}(a)}^{M-1} \delta\right)_{m\,i}^{k\,j} +
\frac{1}{M^2} \, \sum_{a\neq b =1}^M {A_{k}}^m(a) \, {A_{n}}^l(b)
\left(\prod_{{\rm graph}(a,b)}^{M-2}
\delta\right)_{l\,m\,i}^{n\,k\,j} \right. \nonumber \\ \left. +
{\cal O}\left(\frac{1}{M^3}\right) \right] = {\delta_i}^j +
{A_{i}}^j + \frac{1}{2!} \, {\left(\hat{A}^2\right)_i}^j + {\cal
O} \left( \frac{1}{3!}\right) \label{exp} \eqa where the case of
interest for us is when ${A_i}^j = {A_{\mu \, i}}^j (x) \, \Delta
x^\mu$ but in this formula we consider $A$ as a constant matrix.
In this expression the limit is taken over a sequence of open,
connected graphs (broken lines) with $M$ wedges\footnote{Please do
not confuse these graphs with the ones which approximate the curve
$\gamma_{xy}$.}; to each wedge we assign $(1 + \hat{A}/M)$ and
glue them via contraction of lower and upper indices; in the
second term of the third line we sum over all possible insertions
(enumerated by $a$) of one $\hat{A}$ into the graph; $A(a)$ means
just the matrix $A$ placed in the $a$--th wedge; graph$(a)$ is a
disconnected broken line
--- the original graph without the $a$--th wedge; in the third
term of the third line we sum over all possible insertions of the
couple of $\hat{A}$'s into the graph; graph$(a,b)$ is a
disconnected broken line --- the original graph without the
$a$--th and $b$--th wedges. The products of $\delta$'s in \eq{exp}
are taken over these graphs in the obvious way: Via contraction of
lower and upper indices.

It is this definition of the exponent which makes the path
ordering (\ref{hol}) triangulation independent in the limit
$L\to\infty$. In fact, the matrix ${\delta_i}^j$ solves the
equation

\begin{figure}
\begin{center}
\includegraphics[scale=0.5]{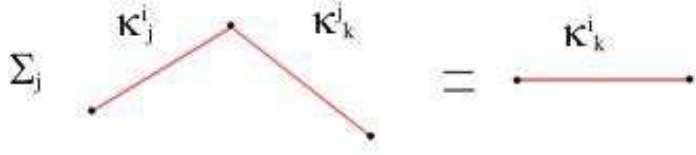}\caption{\footnotesize
The one--dimensional condition for the triangulation
independence.}\label{fig2}
\end{center}
\end{figure}

\bqa {\kappa_i}^j \, {\kappa_j}^k = {\kappa_i}^k,\label{1} \eqa
graphically represented in fig. \ref{fig2}. As the result all
products of $\delta$'s in \eq{exp} are equal to $\delta$ itself:
$\prod \delta = \delta$.

Any solution of the \eq{1} is suitable to define the exponent of a
matrix with two indices and, hence, the triangulation independent
path ordering. The generic solution to this equation is just a
projection operator. Hence, the definition of the exponent with
such a projector is just a standard one when restricted to the
eigen--space of the projector. Using \eq{1} we obviously obtain
that in the limit in question:

\bqa {U\left(\gamma_{xy}\right)_i}^j = {\left(P e^{\int_0^{2\pi} d
s \, \hat{A}_\mu(x) \,
\dot{x}^\mu(s)}\right)_i}^j \equiv \nonumber \\
\equiv {\delta_i}^j + \int_0^{2\pi} ds \, {A_{\mu \, i}}^j(x) \,
\dot{x}^\mu(s) + \int_0^{2\pi} d s_1 \int_0^{s_1} d s_2 \,
{A_{\mu\, i}}^{k}(x) \, \dot{x}^\mu(s_1)\, {A_{\nu\, k}}^j(x) \,
\dot{x}^\nu(s_2)  + \dots, \label{pathor}\eqa where $x(s)$ is the
map (of $s \in [0,2\pi)$ into $X$) whose image is the curve
$\gamma_{xy}$; $x(0) = x, \,\, x(2\pi) = y$.

\begin{figure}
\begin{center}
\includegraphics[scale=0.5]{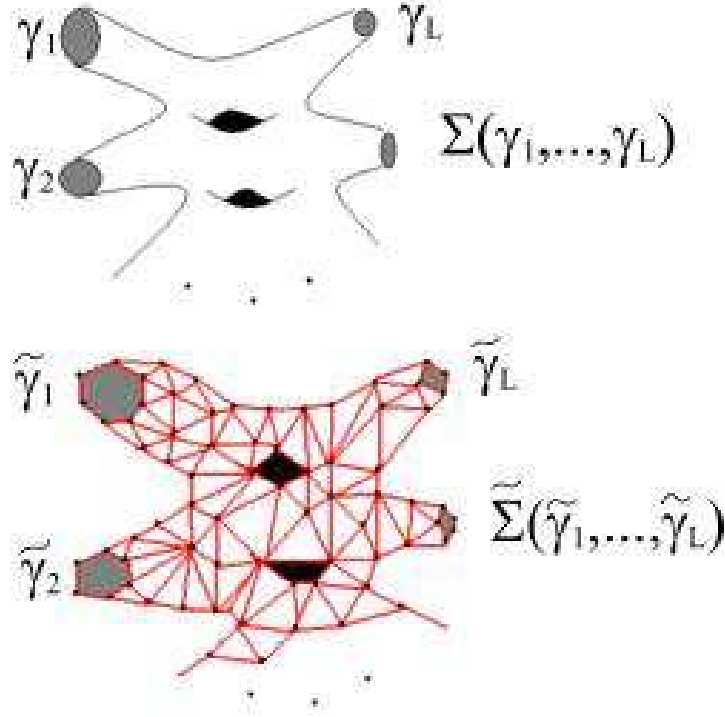}\caption{\footnotesize $\tilde{\Sigma}$
is the discretization of $\Sigma$; $\tilde{\gamma}$'s are
discretizations of $\gamma$'s.}\label{fig3}
\end{center}
\end{figure}

We would like to generalize this construction to the case of the
ordering over two--dimensional surfaces. It is natural to
consider, within this context, a triangulated approximation
$\tilde{\Sigma}(\tilde{\gamma_1}, \dots, \tilde{\gamma_L})$ of an
oriented Riemann surface $\Sigma(\gamma_1, \dots, \gamma_L)$ with
$L$ boundary closed loops $\gamma$'s which are approximated by the
closed broken lines $\tilde{\gamma}$'s (see fig. \ref{fig3}).

Similarly to the one--dimensional case, in which we assign
${U_i}^j$ to each simplex (wedge), in this two--dimensional case
we assign $U_{ijk}$ to each simplex (triangle). At the same time
the indices $i, j$ and $k$ are assigned to the three wedges of the
corresponding triangle. Hence, in this case $\hat{U}: V^3 \to \C $
for an $N$--dimensional vector space $V$ --- the ``fiber'' of our
"fiber bundle".

\begin{figure}
\begin{center}
\includegraphics[scale=0.5]{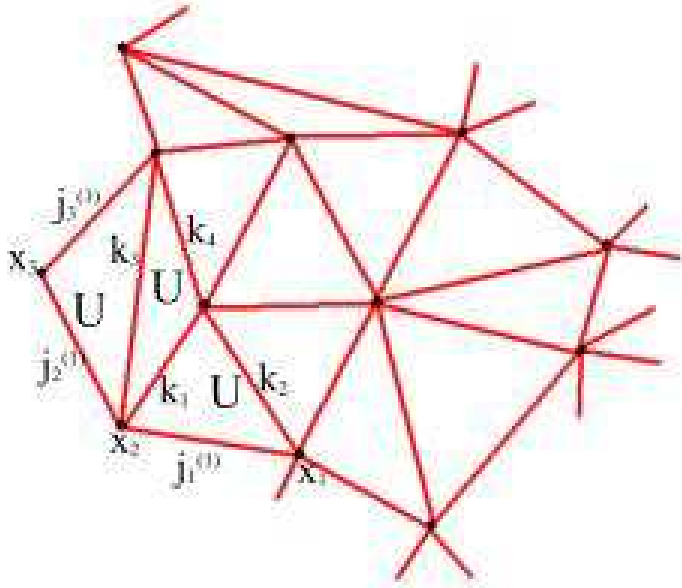}\caption{\footnotesize This is a part of
$\tilde{\Sigma}(\tilde{\gamma}_1,\dots, \tilde{\gamma}_L)$. On
each triangle of this figure there is $U$ matrix with three
indices and we sum over indices assigned to the internal wedges of
the graph.}\label{fig4}
\end{center}
\end{figure}

If we consider cyclicly symmetric $U_{ijk}$ then the
area--ordering prescription is unambiguous. The area--ordering is
obtained by gluing, via the use of a bi--linear form $\kappa^{ij}:
\C \to V^2$, the matrices $U_{ijk}$ on each simplex over the whole
simplicial surface (see fig. \ref{fig4}):

\bqa U\left[\tilde{\Sigma}(\tilde{\gamma_1}, \dots,
\tilde{\gamma_L})\right]_{j^{(1)}_1\dots j^{(1)}_{n_1}\left|
j^{(2)}_1\dots j^{(2)}_{n_2}\right| \dots \left| j^{(L)}_1 \dots
j^{(L)}_{n_L}\right.} \equiv \nonumber \\ \equiv \sum_{k_1, k_2,
\dots, k_w} U_{j^{(1)}_1\, k_1\, k_2}
\left(x_1^{\phantom{\frac12}}, \Delta\sigma_1\right) \, {U^{k_1 \,
k_3}}_{k_4} \left(x_2^{\phantom{\frac12}}, \Delta\sigma_2\right)\,
{U^{k_3}}_{j^{(1)}_2\, j^{(1)}_3} \left(x_3^{\phantom{\frac12}},
\Delta\sigma_3\right)\dots \label{areaod}\eqa where $w$ is the
total number of internal wedges of the graph; $j^{(l)}$ are the
indices corresponding to the wedges of the $l$-th broken line
$\tilde{\gamma}_l$ and $n_l$ is the total number of its wedges,
correspondingly. We higher (lower) the indices via the use of the
aforementioned bilinear form $\kappa^{ij}$ ($\kappa_{ij}
\kappa^{jk} = \delta_i^k$); $\Delta\sigma$'s are elementary
oriented areas associated to the triangles.

\begin{figure}
\begin{center}
\includegraphics[scale=0.3]{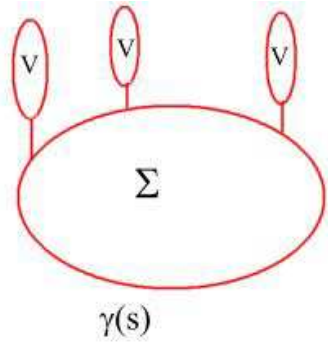}\caption{\footnotesize At each point $s$
of the curve $\gamma$ there is the fiber $V$.}\label{oper}
\end{center}
\end{figure}

  In the continuum limit the discrete indices $1, ..., n_l$ are converted into the
continuum ones $s_l \in [0,\, 2\pi)$ and we obtain:

\bqa U\left[\tilde{\Sigma}(\tilde{\gamma_1}, \dots,
\tilde{\gamma_L})\right]_{j^{(1)}_1\dots j^{(1)}_{n_1}\left|
j^{(2)}_1\dots j^{(2)}_{n_2}\right| \dots \left| j^{(L)}_1 \dots
j^{(L)}_{n_L}\right.} \longrightarrow
U\left[\Sigma^{\phantom{\frac12}}(\gamma_1, \dots,
\gamma_L)\right]_{j^{(1)}(s_1)\left| j^{(2)}(s_2)\right| \dots
\left| j^{(L)}(s_L)\right.}, \nonumber \\ {\rm where} \quad
\hat{U}(\Sigma): V^{\infty}(1) \otimes \dots \otimes V^{\infty}(L)
\to \C, \label{cont}\eqa and $j^{(l)}(s_l)$ is the ``color'' index
assigned to the continuous number of points enumerated by $s_l$
--- a parametrization of the $l$-th loop $\gamma_l$. Such an
operator $\hat{U}(\Sigma)$ for $L=1$ is graphically represented in
fig. \ref{oper}.
The expression (\ref{cont}) is not so unfamiliar for the string
theoreticians as it could seem from the first sight. In fact, if
we substitute $j^{(l)}(s_l)$ by $x^{(l)}(s_l)$, then
$U\left[\Sigma(\gamma_1, \dots, \gamma_L)\right]$ can be
considered as the string amplitude whose end--loops $\gamma$'s are
mapped by $x(s)$'s. This analogy is helpfull in understanding how
our considerations could help in defining the ``two--time''
Hamiltonian formalism --- the formalism for string--like objects.
See short discussion on this subject in the concluding section.

Thus, we have defined the area--ordering prescription. But this is
not the whole story: We have to define $U_{ijk}$ in such a way
that the limit (\ref{cont}) exists and does not dependent on the
triangulation, i.e. does not depend on the way it is taken.

\section{{\it Exponentiation} of a matrix with three indices}

In this section we show that for the continuum expression in
\eq{cont} to be triangulation independent each $U_{ijk}$ should be
represented as the {\it exponent}:

\bqa U_{ijk} = \left(E^{\hat{B}} \right)_{ijk},\label{expo}\eqa
where $\hat{B}$ is a matrix with three indices: particular case of
interest for us is when $B^{ijk}(x, \Delta \sigma) =
B^{ijk}_{\mu\nu}(x) \, \Delta x^\mu\, \Delta x^\nu$. It is natural
to assume that $B$ as well as $U$ has three indices rather than
any other number.

 We take the generalization of (\ref{exp}) to define the exponent of a matrix
with three indices:

\bqa {\rm Tr}\, \left( E^{\hat{B}}_{g,I, \kappa}\right) \equiv
\lim_{M\to\infty} \prod_{{\rm graph}, g}^M \left(\hat{I} +
\frac{\hat{B}}{M}\right),\label{def}\eqa where the limit is taken
over \underline{any} sequence of \underline{closed} (because on
the LHS we take Tr), connected, oriented triangulation
graphs\footnote{Please do not confuse these graphs with the ones
mentioned above, which approximate the surface $\Sigma(\gamma_1,
\dots, \gamma_L)$ in the target space $X$. By ``triangulation
graphs'' we refer to the graphs which triangulate Riemann
surfaces.} of genus $g$ and with $M$ faces (triangles). To take
the limit $M\to\infty$ we have to choose a sequence of graphs.
There is no any distinguished sequence of graphs. Hence, we have
to demand somehow that the result of the limit (\ref{def}) should
not depend on the chosen sequence of graphs.

\begin{figure}
\begin{center}
\includegraphics[scale=0.3]{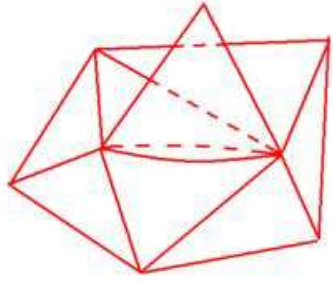}\caption{\footnotesize No such configurations
inside the graphs \underline{defining the
exponent}.}\label{triang}
\end{center}
\end{figure}

We are going to explain now the conditions (including the ones
imposed on the matrices $\kappa^{ij}$ and $I_{ijk}$), under which
the limit (\ref{def}) does not depend on the chosen sequence of
graphs. First, we demand that in these graphs any two triangles
can not meet (glued) at more than one wedge (see fig.
\ref{triang}). Second, as the limit $M\to\infty$ is taken the
number of triangles meeting at any vertex of the graph should be
suppressed in comparison with $M$. This condition is necessary to
overcome the difficulty discussed in the Appendix.

As we show below the definition of the {\it exponent} (\ref{def})
does depend on the choice of matrices $\hat{I}$, $\kappa$ and the
genus $g$ of the graphs in the sequence, along which the limit
$M\to\infty$ is taken. Due to this fact we designate the
definition of the {\it exponent} in \eq{def} with the
corresponding subscripts. At fixed $M$ the product in \eq{def} is
taken over the graph in question: Meaning that at each face of the
graph we put matrix $(I + B/M)_{ijk}$ and we glue indices of these
matrices with the use of the aforementioned bilinear form
$\kappa^{ij}$. The matrices $I_{ijk}$ and $\kappa^{ij}$ we are
going to define in a moment.

\begin{figure}
\begin{center}
\includegraphics[scale=0.5]{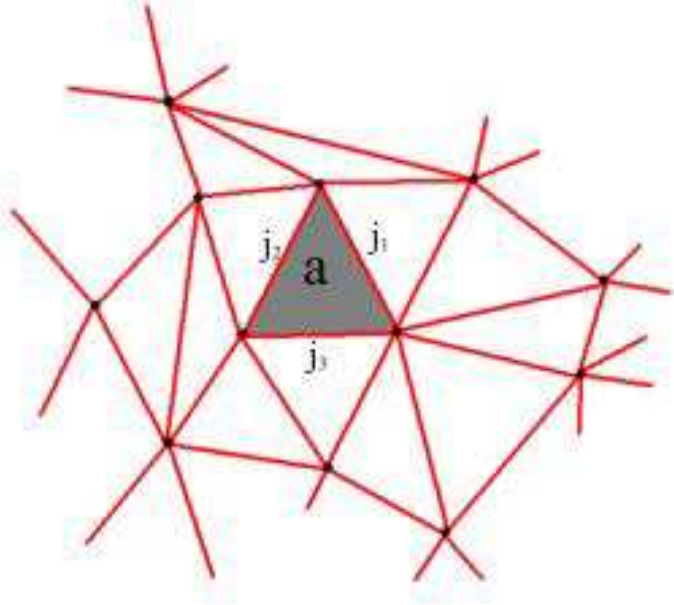}\caption{\footnotesize This is a part of the graph
with the disc topology --- the original spherical graph without
the face $a$.}\label{fig5}
\end{center}
\end{figure}

\begin{figure}
\begin{center}
\includegraphics[scale=0.5]{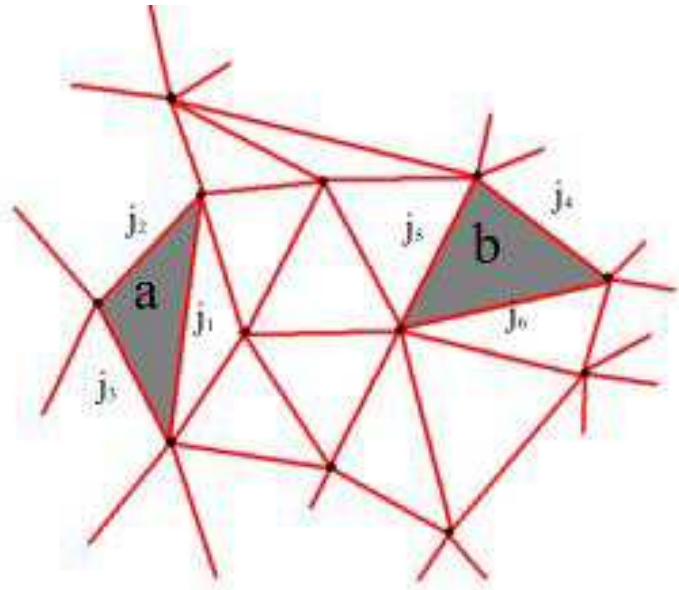}\caption{\footnotesize This is a part of the graph
with the cylindrical topology --- the original spherical graph
without $a$ and $b$ faces.}\label{fig6}
\end{center}
\end{figure}

\begin{figure}
\begin{center}
\includegraphics[scale=0.7]{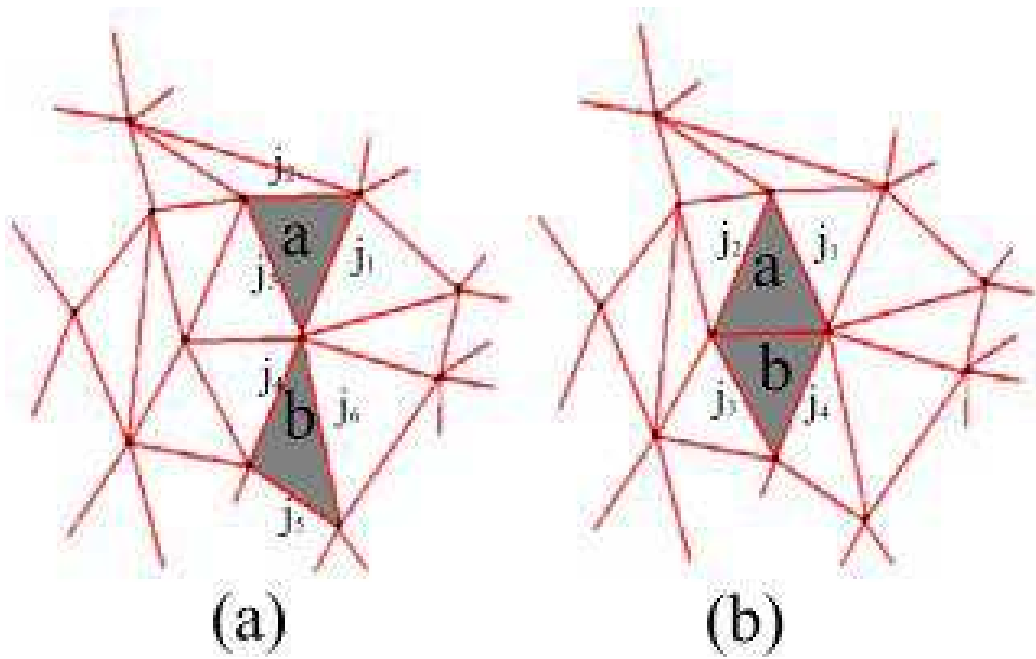}\caption{\footnotesize There are two possibilities
how two triangles can meet each other inside the triangulation
graph.}\label{fig7}
\end{center}
\end{figure}

We start our consideration with an arbitrary oriented graph with
the spherical topology ($g=0$) and $M$ faces: All our
considerations below can be easily generalized to the case of
arbitrary graphs with higher topology. Let us expand the product
in \eq{def} for such a particular choice of the graph:

\bqa\prod_{\rm graph, 0}^M \left(\hat{I} +
\frac{\hat{B}}{M}\right) = \prod_{\rm graph}^M \hat{I} +
\frac{1}{M} \sum_{a=1}^M B^{j_1\,j_2\,j_3}(a) \, \left(\prod_{{\rm
graph}(a)}^{M-1} \hat{I}\right)_{j_3\,j_2\,j_1} + \nonumber \\ +
\frac{1}{M^2} \left[ \sum^{(I)}_{a\neq b}
B^{j_1\,j_2\,j_3}(a)\,B^{j_4\,j_5\,j_6}(b) \, \left(\prod_{{\rm
graph}(I,a,b)}^{M-2} \hat{I}\right)_{j_3\,j_2\,j_1\left| j_6\,
j_5\, j_4\right.} \right. + \nonumber \\ + \sum^{(II)}_{a \neq b}
B^{j_1\,j_2\,j_3}(a)\,B^{j_4\,j_5\,j_6}(b) \, \left(\prod_{{\rm
graph}(II,a,b)}^{M-2} \hat{I}\right)_{j_6 \, j_5\, j_4 \,
j_3\,j_2\, j_1} + \nonumber
\\  \left. +
\sum^{(III)}_{a \neq b} B^{j_1\,j_2\,k}(a)\,{B_k}^{\,j_3\,j_4}(b)
\, \left(\prod_{{\rm graph}(III, a,b)}^{M-2} \hat{I}\right)_{j_4\,
j_3\,j_2\, j_1}\right] + {\cal
O}\left(\frac{1}{M^3}\right),\label{exp1}\eqa where the first term
represents the product of $\hat{I}$'s over the graph and it is a
constant independent of $\hat{B}$. The second term is the sum over
the faces of the graph enumerated by $a$; $\hat{B}(a)$ means the
matrix $\hat{B}$ placed in the face $a$ and the product of
$\hat{I}$'s in this term is taken over the graph$(a)$, which is
the original graph without the face $a$ (see fig. \ref{fig5}). In
the third term the sum ($\sum^{(I)}$) is taken over the remote
faces $a$ and $b$ (see fig. \ref{fig6}). Hence, the graph$(I,a,b)$
(the original graph without $a$ and $b$ faces) has the cylindrical
topology with three external wedges at each end of the cylinder:
This is the reason why we divide the subscript in the
corresponding term into two groups $j_1\,j_2\,j_3$ and
$j_4\,j_5\,j_6$ (we come back to this point later). At the same
time, in the fourth and fifth terms the sums ($\sum^{(II)},
\sum^{(III)}$) are taken over the adjacent faces $a$ and $b$.
Hence, the graphs$(II, III, a,b)$ have disc topology (see fig.
\ref{fig7} (a) and (b)). In the third term the faces $a$ and $b$
are touching each other via one common vertex (see fig. \ref{fig7}
(a)). In the fourth term the faces $a$ and $b$ are touching each
other via one common wedge (see fig. \ref{fig7} (b)). The terms
$\propto B_{i\,k_1\,k_2} \, {B^{k_1\, k_2}}_j$ do not appear in
\eq{exp1} for the reason that we do not consider the graphs having
the configuration of triangles shown in the fig. \ref{triang}.

\begin{figure}
\begin{center}
\includegraphics[scale=0.5]{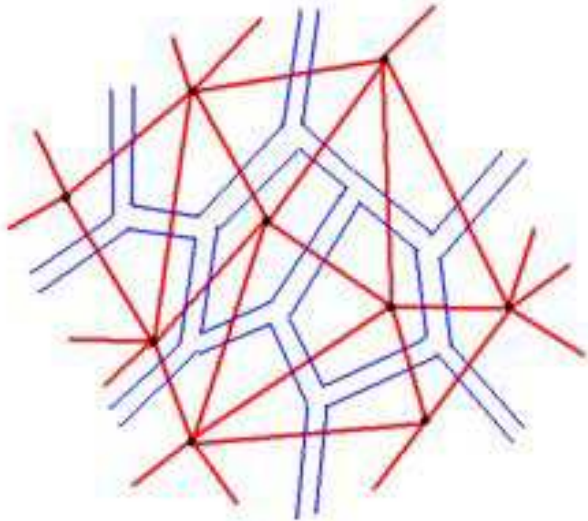}\caption{\footnotesize Duality relation
between graphs. Dual graph is shown here by fat
stripes.}\label{dual}
\end{center}
\end{figure}

\begin{figure}
\begin{center}
\includegraphics[scale=0.5]{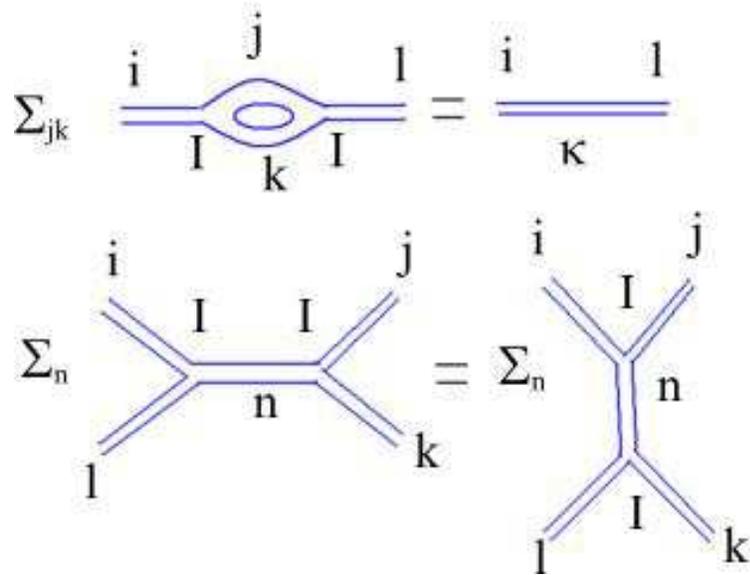}\caption{\footnotesize Two--dimensional
conditions for the triangulation independence.}\label{konditions}
\end{center}
\end{figure}

We would like to choose such matrices $I_{ijk}$ and $\kappa^{ij}$
that the expression (\ref{exp1}) has a well defined limit as
$M\to\infty$. It appears that to reach this goal is the same as to
make the continuum expression (\ref{cont}) independent of
triangulation \cite{Fukuma:1992hy}. It is this moment when we have
to recall \eq{1}. We have to write an analog of this equation for
$\hat{I}$ and $\hat{\kappa}$: So that various triangulations of
the graphs with the same topology will be equal to each other. The
analog of \eq{1} for this case is\footnote{Note that here we
consider only cyclicly symmetric $I_{ijk}$: Then the products over
graphs are defined unambiguously.} $I_{ijk}$:

\bqa\sum_{j,k=1}^N {I_{ij}}^k \, {I_{lk}}^j = \kappa_{il}
\nonumber
\\ \sum_{n=1}^N I_{inl} \, {I^n}_{jk} = \sum_{n=1}^N I_{ijn} \, {I^n}_{kl}.
\label{cond}\eqa The graphical representation of these equations
(in terms of dual three--valent graphs\footnote{From now on we
will be frequently changing from triangulation graphs to the dual
\underline{fat} three--valent ones. The latter can be obtained as
follows. We place the vertices of the dual graph at the centers of
the faces (triangles) of the original one and join the vertices
via the fat wedges of the dual graph passing through the wedges of
the original one (see fig. \ref{dual}). Thus, the dual graph to a
triangulation one is three--valent, i.e. there are three wedges
terminating at each its vertex.}) one can find in the fig.
\ref{konditions}. We will discuss various solution of the
conditions (\ref{cond}) in the next section. Unlike the one
dimensional case, where there is basically the unique solution to
the \eq{1}, in this case we will have the whole zoo of solutions.
Now let us see how these observations influence the definition of
the {\it exponent}.

\begin{figure}
\begin{center}
\includegraphics[scale=0.3]{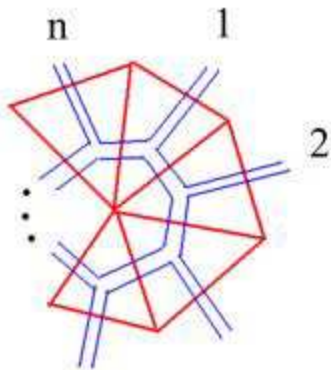}\caption{\footnotesize Disc topology.}\label{ndisc}
\end{center}
\end{figure}

\begin{figure}
\begin{center}
\includegraphics[scale=0.3]{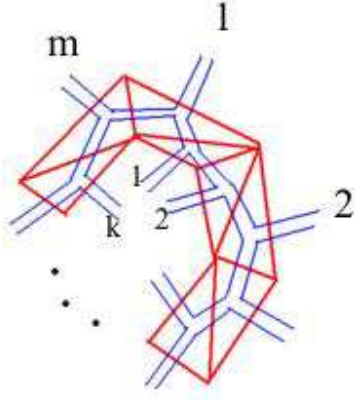}\caption{\footnotesize Annulus topology.
Dual graph is shown for convenience of the understanding of the
symmetries of the corresponding matrix: We can exchange the
positions of say the line number $m$ and $k$ via the second
relation shown in fig. \ref{konditions}. This fact should be
compared with the property of the graph in fig. \ref{ndisc}. There
in general situation we have no means to exchange the positions of
say the $n$--th and the $1$--st external legs.}\label{mkcyl}
\end{center}
\end{figure}

  If these conditions are obeyed, one can map product of $\hat{I}$'s over any graph
to the product of $\hat{I}$'s over any other graph with the same
topology \cite{Fukuma:1992hy} (with the same genera, number of
holes and distribution of external wedges among the holes). This
can be straightforwardly seen by explicit manipulations with the
dual graphs. For example, any graph with the disc topology and $n$
external wedges can be mapped to the graph shown in the fig.
\ref{ndisc}. At the same time, any graph with the annulus topology
and $m$ wedges at one end and $k$ wedges at another end can be
mapped to the graph shown in the fig. \ref{mkcyl}. The expressions
for such two kinds of multiplications of $\hat{I}$'s are not
equivalent to each other even if $n=k+m$. It is such facts which
make us to distinguish the third and the fourth terms in \eq{exp1}
and to fix the genus $g$ of the graphs in the sequence to define
the {\it exponent}.

\begin{figure}
\begin{center}
\includegraphics[scale=0.3]{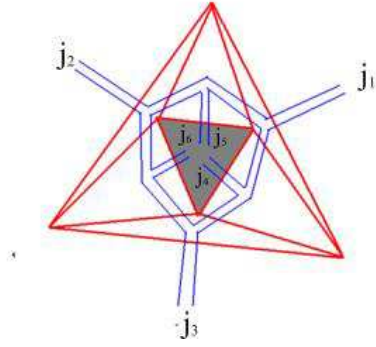}\caption{\footnotesize Graph for
$I_{j_3\,j_2\,j_1| j_6\,j_5\,j_6}$. Dual graph is shown to explain
the symmetry properties of the corresponding matrices.}\label{I33}
\end{center}
\end{figure}

\begin{figure}
\begin{center}
\includegraphics[scale=0.3]{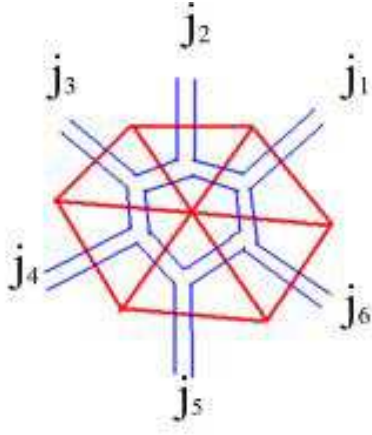}\caption{\footnotesize Graph for
$I_{j_6\,j_5\,j_4\,j_3\,j_2\,j_1}$.}\label{I6}
\end{center}
\end{figure}

 Thus, if $I_{ijk}$ and $\kappa^{ij}$ obey \eq{cond}, we obtain:

\bqa\prod_{\rm graph, 0}^M \left(\hat{I} +
\frac{\hat{B}}{M}\right) = N + B^{j_1\,j_2\,j_3} \,
I_{j_3\,j_2\,j_1} + \nonumber
\\ + \frac{1}{M^2} \left[C_1(M) \, B^{j_1\,j_2\,j_3\phantom{\frac12}}B^{j_4\,j_5\,j_6}\,
I_{j_3\,j_2\,j_1\left| j_6\, j_5\, j_4\right.} \right. + C_2(M) \,
B^{j_1\,j_2\,j_3}\,B^{j_4\,j_5\,j_6} \, I_{j_6 \dots j_1} +
\nonumber \\  \left.  + C_3(M) \,
B^{j_1\,j_2\,k\phantom{\frac12}}{B_k}^{\,j_3\,j_4} \, I_{j_4\dots
j_1} \right] + {\cal O}\left(\frac{1}{M^3}\right),\label{exp2}\eqa
where we denote by the same letter $I$ the product of the
three--linear form $I_{ijk}$ over the corresponding graphs.
$I_{j_3\,j_2\,j_1 | j_6\, j_5\, j_4}$ is given by the annulus
graph in fig. \ref{I33} where in each face stands matrix
$I_{ijk}$. As can be seen explicitly from the dual graph
representation, the matrix $I_{j_3\,j_2\,j_1 | j_6\, j_5\, j_4}$
is cyclicly symmetric in the $j_1, \, j_2, \,j_3$ and $j_4, \,
j_5, \, j_6$ indices separately. But otherwise one can exchange
position of indices from these two groups in any order (see the
discussion under the fig. \ref{mkcyl}). This property should be
contrasted with the one of the matrix $I_{j_6 \dots j_1}$ which is
given by the disc graph in fig. \ref{I6}. The matrix $I_{j_6 \dots
j_1}$ is symmetric under the cyclic exchange of all its six
indices $j_1, \dots, j_6$, but for generic choice of $I_{ijk}$
there is no any other symmetry. The same is true for the matrix
$I_{j_4\dots j_1}$ which is given by the graph with disc topology,
but with four external wedges: It is symmetric under the cyclic
exchange of all its four indices.

\begin{figure}
\begin{center}
\includegraphics[scale=0.5]{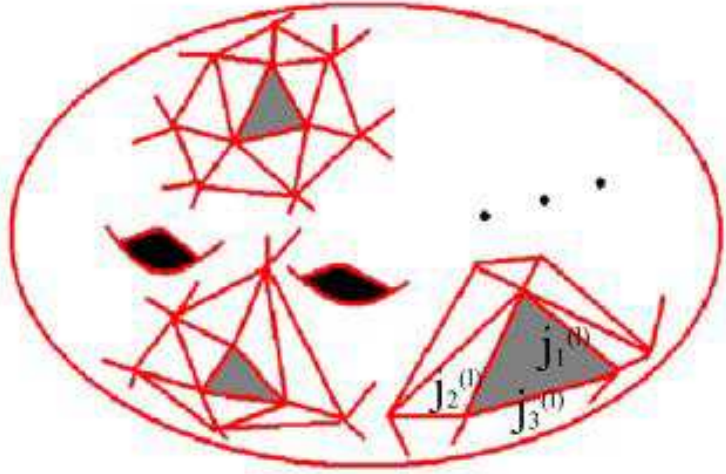}\caption{\footnotesize Triangulation graph
defining $I^g$.}\label{Ig}
\end{center}
\end{figure}

The constants $C_1(M)$, $C_2(M)$ and $C_3(M)$ are the numbers of
the terms in the corresponding sums in \eq{exp1}. We calculate
them in the Appendix and show that under conditions listed below
the \eq{def} the constants $C_2(M)$ and $C_3(M)$ are $\propto M$
and do not survive the limit $M\to\infty$ and

$$\frac{C_1(M)}{M^2} \to \frac{1}{2!}.$$
Similar story happens for the higher terms in $\hat{B}$ and higher
genera (see the Appendix). As the result we obtain the following
expression for the {\it exponent}:

\bqa {\rm Tr} \, \left(E^{\hat{B}}_{g,I,\kappa}\right) =
\sum_{F=0}^{\infty} \frac{1}{F!} \,
B^{j^{(1)}_1\,j^{(1)}_2\,j^{(1)}_3} \dots
B^{j^{(F)}_1\,j^{(F)}_2\,j^{(F)}_3} \,
I^g_{j^{(1)}_3\,j^{(1)}_2\,j^{(1)}_1\left| \dots \right|
j^{(F)}_3\,j^{(F)}_2\,j^{(F)}_1}, \label{defexp}\eqa where
$I^g_{j^{(1)}_3\,j^{(1)}_2\,j^{(1)}_1\left| \dots \right|
j^{(F)}_3\,j^{(F)}_2\,j^{(F)}_1}$ is the matrix with $3F$ indices
obtained via multiplication of $I_{ijk}$ over \underline{any}
oriented graph with $g$ handles, $F$ holes and 3 external wedges
at each hole (see fig. \ref{Ig}). At the same time $I^g, (F=0)$ is
just a number obtained by the multiplication of $\hat{I}$ over the
closed genus $g$ graph. For example, $I^0 = \sum_i \delta_i^i = N$
because of the first relation in \eq{cond}.

Similarly one can obtain the definition of the {\it exponent} for
an open graph with any number of external wedges:

\bqa \left(E^{\hat{B}}_{g,I,\kappa}\right)_{\left. m^{(1)}_1\dots
\, m^{(1)}_{n_1}\right| \dots \left|m^{(L)}_1\dots \,
m^{(L)}_{n_L}\right.} = \nonumber \\ = \sum_{F=0}^{\infty}
\frac{1}{F!} \, B^{j^{(1)}_1\,j^{(1)}_2\,j^{(1)}_3} \dots
B^{j^{(F)}_1\,j^{(F)}_2\,j^{(F)}_3} \, I^g_{\left. m^{(1)}_1\dots
\, m^{(1)}_{n_1}\right| \dots \left|m^{(L)}_1\dots \,m^{(L)}_{n_L}
\right| j^{(1)}_3\,j^{(1)}_2\,j^{(1)}_1\left| \dots \right|
j^{(F)}_3\,j^{(F)}_2\,j^{(F)}_1}, \eqa where $I^g_{\left.
m^{(1)}_1\dots \, m^{(1)}_{n_1}\right| \dots \left|m^{(L)}_1\dots
\, m^{(L)}_{n_L} \right| j^{(1)}_3\,j^{(1)}_2\,j^{(1)}_1\left|
\dots \right| j^{(F)}_3\,j^{(F)}_2\,j^{(F)}_1}$ is the matrix with
$3F + n_1 + \dots + n_L$ indices obtained via multiplication of
$\hat{I}$'s over \underline{any} oriented graph with $g$ handles,
$L+F$ holes, 3 external wedges at $F$ holes and $n_l$ external
wedges at $l$--th hole ($l = 1,\dots , L$).

\section{Properties of the area--ordering and other definitions of the
{\it exponent}}

Let us see now what kind of the area--ordering we obtain with such
a definition of the exponent. The area--ordered {\it exponent}
(``$A\,E$'') of the non--Abelian $B$--field over the disc $D$ is
equal to:

\bqa U_{j(s)}(D, \hat{B}) = \left(A \,
E_{g,I,\kappa}^{{\int\int}_{D} \hat{B}_{\mu\nu} \, dx_\mu \,
dx_\nu}\right)_{j(s)} \equiv \sum_{F=0}^{\infty} \frac{1}{F!} \,
I^g_{\left. j(s)\right| j^{(1)}_3\,j^{(1)}_2\,j^{(1)}_1\left|
\dots \right| j^{(F)}_3\,j^{(F)}_2\,j^{(F)}_1} \, \times \nonumber
\\ \times {\int\int}_{D} B_{\mu_1\nu_1}^{j^{(1)}_1\,j^{(1)}_2\,j^{(1)}_3}
dx_{\mu_1} \, dx_{\nu_1} \dots {\int\int}_{D}
B_{\mu_F\nu_F}^{j^{(F)}_1\,j^{(F)}_2\,j^{(F)}_3} dx_{\mu_F} \,
dx_{\nu_F},\label{areaod1} \eqa where $s$ is the parametrization
of the boundary and $j(s)$ is the index function at the boundary
$\pr D$. Frankly speaking we are not sure whether the product of
$I$'s leading to $I^g_{j(s)|\dots}$ with continuous index can be
made perfectly meaningful in the case when $j$ takes discrete
values. This demands a separate careful study. For our
considerations in this section we can regularize \eq{areaod1} to
convert $s$ into a discrete set.

In any case, such an area--ordering is rather trivial because
there is no need to order anything. In fact, due to the specific
features (mentioned above for the case of $I_{j_3\,j_2\,j_1 |
j_6\, j_5\, j_4}$) of the matrix $I^g_{\left. j(s)\right|
j^{(1)}_3\,j^{(1)}_2\,j^{(1)}_1\left| \dots \right|
j^{(F)}_3\,j^{(F)}_2\,j^{(F)}_1}$, we can easily interchange the
order of integrals over $B$ in \eq{areaod1}. As the result the
non--Abelian tensor field $\hat{B}$ is just a collection of $N^3$
Abelian fields $B$ enumerated by three indices $i$, $j$ and $k$:
The gauge transformations\footnote{We will discuss the gauge
transformations for such objects as (\ref{areaod1}) in a separate
publication.} and the field strength can be found to be the same
as for the ordinary string two--tensor $B$--field for every triple
$i,j,k$. In fact, all the non--trivial (contact) terms in
\eq{exp2} have vanished with such constants as $C_2(M)$ and
$C_3(M)$.

However, this does not mean that we have obtained a trivial
bundle! In fact, if we consider a surface $\Sigma$ with one
designated point $x$ and two external wedges at this point the
corresponding $U(\Sigma_x)_{ij}$ (or ${U(\Sigma_x)_i}^j =
U(\Sigma_x)_{ik} \, \kappa^{kj}$) gives a nontrivial non--Abelian
map. This is drastically different from the case when the ordinary
gauge connection $\hat{A}_\mu$ becomes a collection of $N$ Abelian
gauge fields. All that goes without saying that more complicated
${U(\Sigma_x)_{ijm\dots}}^{kl\dots}$ give completely non--trivial
non--linear maps.

The only way to define more complicated area--ordering is to
define the {\it exponent} in a different way (if possible at all):
So that such terms as $C_2$ and $C_3$ survive the limit
$M\to\infty$. For example, we can  abandon the second condition
presented after the \eq{def} and take the diamond type graphs (see
fig. \ref{sphere}) to define the {\it exponent}. Then, the $C_2$
terms will survive and we will obtain the non--trivial {\it
exponent} (see the Appendix):

\bqa {\rm Tr}\, \left(E^{\hat{B}}_{g,I,\kappa, K}\right) = I^g +
B^{j_1\,j_2\,j_3} \, I^g_{j_3\,j_2\,j_1} + \frac{1}{2!}
\left[\left(1 - \frac{1}{2(K+1)^2}\right) \,
B^{j_1\,j_2\,j_3\phantom{\frac12}}B^{j_4\,j_5\,j_6}\,
I^g_{j_3\,j_2\,j_1\left| j_6\, j_5\, j_4\right.} + \right. \nonumber \\
\left. + \frac{1}{2(K+1)^2} \,
B^{j_1\,j_2\,j_3}\,B^{j_4\,j_5\,j_6} \, I^g_{j_6 \dots j_1}
\right] + {\cal O}\left(\frac{1}{3!}\right) \quad {\rm where}
\quad K\in Z_+.\eqa But the definition of the {\it exponent} in
such a way seems to us as the abuse on the nature. We would like
to see a more natural definition of a non--trivial {\it exponent}:
More similar to the one presented in the previous section.

At this stage we can propose the only other possibility for the
definitions of the {\it exponent}. It is inspired by \eq{def}. We
can put in the faces of the graphs in \eq{def} $I_{ijk}$ matrix
rather than $I_{ijk} + B_{ijk}/M$, but glue them with the use of
$\kappa^{ij} + B^{ij}/M$ rather than just with --- $\kappa^{ij}$,
i.e.:

\bqa{\rm Tr}\, \left( \tilde{E}^{\hat{B}}_{g,I,\kappa}\right)
\equiv \lim_{L\to\infty} \prod_{{\rm graph}(g)}^L
\left(\hat{\kappa} + \frac{\hat{B}}{L}\right),\eqa where ``graph''
is a triangulation graph of genus $g$ with $L$ wedges; $B$ in this
case obviously has two indices rather than three. In the light of
the above discussion this definition leads to the triangulation
independent area--ordering. The result for the {\it exponent} is:

\bqa {\rm Tr} \, \left(\tilde{E}^{\hat{B}}_{g,I,\kappa}\right) =
\sum_{F=0}^{\infty} \frac{1}{F!} \, B^{j^{(1)}_1\,j^{(1)}_2} \dots
B^{j^{(F)}_1\,j^{(F)}_2} \, I^g_{j^{(1)}_1\,j^{(1)}_2\left| \dots
\right| j^{(F)}_1\,j^{(F)}_2},\eqa where
$I^g_{j^{(1)}_1\,j^{(1)}_2\left| \dots \right|
j^{(F)}_1\,j^{(F)}_2}$ is the matrix with $2F$ indices obtained
via multiplication of $I_{ijk}$ over \underline{any} oriented
graph with $g$ handles, $F$ holes and 2 external wedges at each
hole.

However, we think (but can not prove) that this new definition of
the {\it exponent} is related to (\ref{def}) by a gauge
transformation:

\bqa g^3\, \left(\hat{I}' + \frac{\hat{B}'}{M}\right) = \hat{I}
\quad {\rm and} \quad \hat{\kappa}' \, g^{-2} = \hat{\kappa} +
\frac{\hat{B}}{M},\eqa where $\kappa'$ and $I'$ obey \eq{cond} as
well and $\hat{B}'$ has three indices. We will discuss such gauge
transformations in a separate publication.

\section{Explicit examples of the $\hat{I}$ and $\hat{\kappa}$ matrices}

Let us discuss some explicit solutions to the \eq{cond}. For
example, the simplest choice is $I_{ijk} = \delta_{ijk}$,
$\kappa^{ij} = \delta^{ij}$, where $\delta_{ijk}$ is represented
by a cubic matrix whose only non-zero elements are units standing
on the main diagonal of the cube. With such a choice of $\hat{I}$
and $\hat{\kappa}$ we obtain that:

$$
I^g_{\left.j^{(1)}_1\dots j^{(1)}_{n_1}\right| \dots
\left|j^{(L)}_1\dots j^{(L)}_{n_L}\right.} =
\delta_{j^{(1)}_1\dots j^{(1)}_{n_1} \dots j^{(L)}_{1} \dots
j^{(L)}_{n_L}},
$$ where $\delta_{j^{(1)}_1\dots j^{(1)}_{n_1} \dots j^{(L)}_{1} \dots
j^{(L)}_{n_L}}$ is $n = n_1 + \dots + n_L$ dimensional cubic
matrix whose only non--zero elements are units standing on its
main diagonal. In fact, if say any two of $j$'s in
$I^g_{\left.j^{(1)}_1\dots j^{(1)}_{n_1}\right| \dots
\left|j^{(L)}_1\dots j^{(L)}_{n_L}\right.}$ are not equal to each
other then in one of the faces inside the graph, defining this
matrix, there are two indices which do not coincide. This means
that corresponding $\delta_{ijk}$ sitting in this face vanishes,
hence, $I^g$ vanishes. Then, for $I^g$ to be non--zero all its
$j$'s have to be equal to each other.

As the result, for such a choice of $I_{ijk}$ and $\kappa_{ij}$
the definition of the {\it exponent} looks as follows:

\bqa{\rm Tr} \,
\left(E^{\hat{B}}_{\delta_{(3)},\delta_{(2)}}\right) =
\sum_{L=0}^{\infty} \frac{1}{L!} \, \sum_j
\left(B^{j\,j\,j}\right)^L = \sum_j e^{B^{j\,j\,j}},\eqa i.e. does
not depend on the genus $g$. Such an {\it exponent} is trivial
because it depends only on the diagonal elements of the matrix
$\hat{B}$.

Another simple choice is when $\kappa^{ij} = \delta^{ij}$ and
$I_{ijk}$ is a diagonal cubic matrix with ether $+1$ or $-1$ at
each place at the diagonal: All the possible distributions of $\pm
1$ are allowed. Then as well we will obtain a trivial {\it
exponent} depending only on the diagonal elements of $B$ (with
alternating signs). Moreover, if one takes diagonal matrices
$I_{ijk}$ and $\kappa^{ij}$ it is easy to see that the
corresponding $\hat{I}^g$'s (for any graphs and with any
combinations of external legs) are all diagonal and, hence, the
corresponding {\it exponent} depends only on the diagonal elements
of $\hat{B}$. Finally, if $I_{ijk}$ is symmetric under any
exchange of its indices then as well all $\hat{I}^g$'s for any
graphs and with any external legs are symmetric under any exchange
of their external indices and does not depend on $g$.

To obtain less trivial {\it exponents} one has to recall that the
conditions (\ref{cond}) are related to the fusion rules
\cite{Fukuma:1992hy}, \cite{Verlinde:1988sn}. The easiest way to
explain these fusion rules is to consider a group\footnote{In
general any associative and semi--simple algebra is suitable to
define the fusion rules.} whose \underline{all} elements are
$\varphi_i$, $i=\overline{1,N}$ (for non--finite groups $i$ takes
continuous values). Consider the product in this group:

\bqa\varphi_i \circ \varphi_j = \sum_k {I_{ij}}^k
\varphi_k,\label{prod}\eqa where the matrix ${I_{ij}}^k$ consists
of $N^2$ non--zero elements spread over the cubic matrix, with all
other elements equal to zero. We choose $\kappa_{ij} = {I_{il}}^k
\, {I_{kj}}^l$ to be the metric on this group. Then $I_{ijk} =
{I_{ij}}^l \, \kappa_{lk}$ is always cyclicly symmetric. The
condition of the associativity for the product (\ref{prod}) is
just the second condition in \eq{cond}.  This gives us the recipe
to construct possible explicit doubles $I_{ijk}$, $\kappa^{ij}$.

As the non--trivial but symmetric in all three indices example one
can consider Abelian $Z_N$ group. Then ${I_{ij}}^k$ has non--zero
elements only if $k = i+j \,\,{\rm mod} \,\, N$. To get
$\kappa_{ij} = \delta_{ij}$ one has to take these non--zero
elements to be equal to $1/\sqrt{N}$. It is straightforward to see
that $I_{ijk}$ is symmetric under any exchange of its indices. For
example, the explicit formula for $\hat{I}$ for the case $N=2$ is
as follows ($i=0,1$ and $\varphi_0 = 1$):

\bqa
I_{000} = I_{110} = I_{101} = I_{011} = \frac{1}{\sqrt{2}} \nonumber \\
I_{111} = I_{001} = I_{010} = I_{100} = 0. \eqa With such a choice
of $\hat{I}$ and $\hat{\kappa}$ the {\it exponent} (\ref{defexp})
depends on non--diagonal values of the matrix $\hat{B}$. However,
it depends only on the symmetric part of the matrix $\hat{B}$.

To obtain non--symmetric $\hat{I}$ with finite number of indices
one can consider any finite non--Abelian group, say $S_n$
--- the group of permutations of $n$ elements ($n!=N$). Once one understood
the idea of finding $\hat{I}$ and $\kappa$ along the way presented
in this section, it is easy to work out the explicit value for
$I_{ijk}$ and $\kappa_{ij}$ for the case of $S_n$ or any other
group. Their explicit values are not relevant for the
consideration in this note. The computation of the explicit values
of the matrices $\hat{I}^g$'s with various distributions of
indexes is technically somewhat complicated exercise, but the
algorithm is obvious and computer will do this job easily. Note
that these matrices are related to the two--dimensional
topological invariants \cite{Fukuma:1992hy}.

\section{Conclusions, Future directions and Acknowledgments}

Thus, we have obtained an explicit area--ordering prescription
which gives a non--trivial bundles on loop spaces. As can be
easily seen this construction reduces when $N=1$ to the ordinary
string two--tensor $B$--field. It is interesting to observe how
the non--Abelian $B$--fields reduce to the ordinary non--Abelian
gauge connections when the surface $\Sigma$  with cylindrical
topology (and one external index at each border of the cylinder)
degenerates into a curve $\gamma$. We will discuss such a
situation in a separate publication.

As well our considerations can be easily generalized to higher
dimensions. In that case triangles are substituted by
higher--dimensional simplicies. Then the corresponding tensor
fields have to carry four, five and etc. indices. We have to
somehow {\it exponentiate} such tensor fields. The definitions of
the corresponding {\it exponents} uses the multi--index
generalizations of the $\hat{I}$ matrices. As well the conditions
(\ref{cond}) are exchanged for the ones following from the
multi--dimensional Matveev or Alexander moves (see e.g.
\cite{Turaev:1992hq}).

It is interesting as well to consider how our considerations are
related to the renormalization group in QFT
\cite{Gerasimov:2000pr}.

And last but not least. Probably the most important feature of the
traditional exponent is that it solves the simple differential
equation. Within this context it is natural to ask which
differential equation is solved by the {\it exponent} defined in
this note? This question is naturally related to the Hamiltonian
dynamics of string--like objects. In fact, consider a quantum
mechanical amplitude:

$$\left\langle x\left| e^{-T \, H}\right| y \right\rangle.$$ This
amplitude solves the obvious differential equation describing
quantum Hamiltonian evolution of the system. The amplitude can be
considered as the holonomy for the connection $H$
--- Hamiltonian. This is the connection on the fiber bundle whose
base is the world line --- the space of $T$ --- rather than the
target space. The fiber is the Hilbert space of the quantum
mechanical system in question. From this point of view $x$ and $y$
play the role of indices.

Similarly we can represent the three--point string amplitude as:

$$ E^{- {\int\int}_{\Sigma} {\cal H} d\sigma\, d\tau}_{0,\hat{I},\hat{\kappa}}
|x\rangle\, |y \rangle \,| z\rangle,$$ where now ${\cal H}$ is the
two--tensor Hamiltonian with three color indices ($x$, $y$ and
$z$) to be presented in a separate publication (along the lines of
\cite{Akhmedov:2004yb}, \cite{Akhmedov:2005mr}). Here the base of
the bundle is the world--sheet --- the space of $\sigma$ and
$\tau$
--- rather than the target space. The fiber is the Hilbert space
as well. It is interesting to see which differential (variational)
equation is satisfied by such an amplitude. The latter equation
will define the ``two--time'' Hamiltonian evolution.

Why such an approach is better than well established ones? We hope
that our approach will allow to address the String Field Theory
directly through the ``$\Phi^3$'' action rather than through
perturbative expansion around ``$\Phi \, Q \, \Phi$''. If this
will work, such an approach will be generalizable for the case of
higher ``Brane Field Theories'' and will lead to a background
independent description of the theories: various $B$ (${\cal H}$)
fields will lead to various backgrounds. The $B$ fields, in their
own right, will be solutions of the equations of motion in the
non--Abelian tensor theory.

 I would like to acknowledge valuable
discussions with A.Morozov, A.Zabrodin, M.Zubkov, F.Gubarev,
T.Pilling, A.Skirzewski, N.Amburg, D.Vasiliev, A.Losev, A.Rosly,
L.Andersen, H.Nicolai, S.Theisen and especially to I.Runkel and
G.Sharigin. I would like to thank V.Dolotin and A.Gerasimov for
very intensive, deep and useful discussions. I would like to thank
H.Nicolai and S.Theisen for the hospitality at MPI, Golm where
this work was finished. This work was done under the partial
support of grants RFBR 04-02-16880, INTAS 03--51--5460 and the
Grant from the President of Russian Federation MK--2097.2004.2.

\section{Appendix}

In this appendix we calculate the constants $C_1(M)$, $C_2(M)$ and
$C_3(M)$. The sum of these constants is obviously given by the
choice of two faces out of $M$:

$$C_1(M) + C_2(M) + C_3(M) = \C^2_M,$$ where $\C^2_M = M(M-1)/2$ is the binomial
coefficient. At the same time $C_3(M)$ is equal to the number of
wedges in the graph (number of two adjacent faces in the graph):

$$C_3(M) = \frac32 \, \C^1_M = \frac32 \, M.$$ The explanation of the
coefficient $\frac32$ is as follows. For the triangulation graphs
there is an equality:  $2 \times ({\rm No. \,\, of\,\, wedges}) =
3 \times ({\rm No.\,\, of \,\, faces})$. Thus, in the limit $M\to
\infty$ we obtain that:

$$\frac{C_3(M)}{M^2}\to 0.$$
As the result the last term in \eq{exp2} does not survive in the
limit.

To find $C_2(M)$ we have to calculate at each vertex the number of
possibilities for two triangles to meet each other in the way
shown in fig. \ref{fig7} (a) and then sum over all vertices. The
number of couples of triangles at each vertex is $\C_{n_v}^2$
where $n_v$ is the total number of triangles meeting at the
$v$--th vertex. From this we have to subtract the possibilities
for the triangles to meet in the way shown in the fig. \ref{fig7}
(b). The number of the latter is given by $l_v$ --- the number
wedges meeting at the $v$--th vertex. In summing over the vertices
we take each wedge into account twice. As the result:

$$C_2(M) = \sum_{v=1}^V \left(\C_{n_v}^2 - \frac{l_v}{2}\right),$$
where $V = M/2 - 2 + 2g$ is the total number of vertices of the
graph. Note that $n_v = l_v$ for any $v$ in a closed graph.
Furthermore, $\sum_v l_v = 2 \, L$, where $L= 3M/2$ is the total
number of wedges of the graph. Hence,

$$C_2(M) = \sum_{v=1}^V \frac{n_v^2}{2} - 2\, L$$
Unfortunately we can not express this value through the number of
faces $M$ and genus $g$ of the graph. However, we can do an
explicit calculation for the regular dual to triangulation graphs.
For example, for the soccer ball graphs (12 pentagons, or 8
squares, or 4 triangles diluted symmetrically into the hexagon
lattice) we have:

$$C_2(M) = 12 \, \frac{5^2}{2} + \left(\frac{M}{2}-2-12\right) \,
\frac{6^2}{2} - 3 \, M$$ for the ``12 pentagon'' case. Hence,

$$\frac{C_2(M)}{M^2}\to 0$$ as $M\to\infty$.
More generally, if the numbers $n_v$ are limited from above by
$n_{max}$ as $M\to\infty$ then $C_2(M)< \frac{n_{max}^2}{4} \, M$
as $M\to\infty$ and, hence, $C_2(M)/M^2 \to 0$.

\begin{figure}
\begin{center}
\includegraphics[scale=0.3]{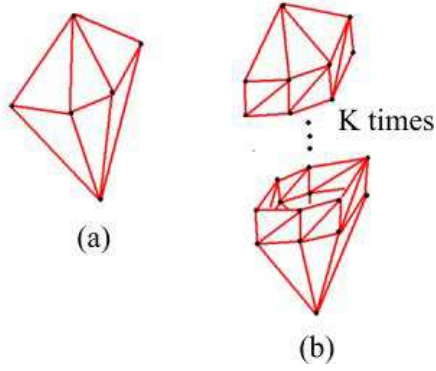}\caption{\footnotesize Diamond type graphs.
In fig. (b) there are $K$ layers of the sequences of rectangles
constructed from couples of triangles.}\label{sphere}
\end{center}
\end{figure}

Let us consider the case when $n_v$ is not limited from above.
Consider the diamond type graphs as in fig. \ref{sphere}(a). In
this case the limit $M\to\infty$ is taken by symmetrically
increasing the number of triangles meeting at the lower and upper
vertices.
For such graphs:

$$C_2(M) = 2\, \frac{(M/2)^2}{2} + M \, \frac{4^2}{2} - 3 \, M$$
and hence $C_2(M)/M^2 \to 1/4$ in the limit $M\to \infty$ and the
$C_2$ term survives the limit. These considerations mean that the
limit (\ref{def}) depends on actual choices of the graphs in the
sequence.

However, if we consider the graphs of type \ref{sphere} (b) with
fixed $K$ then:

$$C_2(M) = \frac{2}{2}\, \left(\frac{M}{2(K+1)}\right)^2 + \frac{M}{K+1} \frac{5^2}{2}
+ \left(\frac{M}{2}-2 - \frac{M}{K+1}\right)\, \frac{6^2}{2} -
3\,M.$$ Hence, if $M\to\infty$ and $K$ is fixed then

$$\frac{C_2(M)}{M^2} \to \frac{1}{4(K+1)^2}$$ and again the $C_2$
term survives the limit. Note, however, that for fixed $K$ the
number of triangles meeting at the poles of the diamond
\ref{sphere} (b) is of the order of $M$.

Now, if we take the limit $K\to\infty$ then $C_2(M)/M^2 \to 0$ and
the number of triangles meeting at the poles of the diamond is
suppressed in comparison with respect to $M$. Hence, to avoid the
complication with the dependence of the limit $M\to\infty$ on the
choice of graphs in \eq{def} we have to take such graphs in which
the number of triangles meeting at each vertex is suppressed in
comparison with $M$. Then $C_2(M)$ does not survive the limit
$M\to\infty$ and the \eq{def} does not depend on the choices of
the graphs in the sequence. Then $C_1(M)/M^2 \to \C_M^2/M^2 \to
1/2!$.

Presented here considerations are valid for graphs of any genera,
i.e. $C_2$ and $C_3$ terms vanish for any genera. Similar but a
little more tedious calculation we performed for the $\hat{B}^3$
terms. It appears that the only surviving term is (if
$n_{max}(M)/M \to 0$ as $M\to\infty$):

$$\frac{1}{3!}\, B^{j^{(1)}_1\,j^{(1)}_2\,j^{(1)}_3} \,
B^{j^{(2)}_1\,j^{(2)}_2\,j^{(2)}_3}
B^{j^{(3)}_1\,j^{(3)}_2\,j^{(3)}_3} \,
I^g_{j^{(1)}_3\,j^{(1)}_2\,j^{(1)}_1\left|
j^{(2)}_3\,j^{(2)}_2\,j^{(2)}_1 \right|
j^{(3)}_3\,j^{(3)}_2\,j^{(3)}_1}.$$ As well we expect similar
story to happen for higher orders in powers of $\hat{B}$. In fact,
consider $F$ matrices $\hat{B}$ spread over a graph with $M$
faces. Then on general grounds we can expect that if $M\to\infty$
configurations where $\hat{B}$'s meet each other are suppressed
(under the conditions listed below \eq{def}) in comparison with
the one where they are separated. As the result we obtain
\eq{defexp}.

\end{document}